# Approximate Solution of Bohr-Mottelson Hamiltonian with Minimal Length Effect for Hulthen Potential Using Asymptotic Iteration Method

**Isnaini Lilis Elviyanti,[1] A Suparmi,[1,2] and C Cari[1,2]**

[1]Physics Department, Graduate Program, Sebelas Maret University, Jl. Ir. Sutami 36A Kentingan, Surakarta 57126, Indonesia

[2]Physics Department, Faculty of Mathematics and Fundamental Science, Sebelas Maret University, Jl. Ir. Sutami 36A Kentingan, Surakarta 57126, Indonesia

Email: isna.elviyanti@gmail.com

**Abstract**. The approximate solution of Bohr-Mottelson Hamiltonian in rigid deformed nucleus case for Hulthen potential with minimal length effect was investigated using Asymptotic Iteration Method. Asymptotic Iteration Method was used to solve approximately the Bohr-Mottelson Hamiltonian to obtain energy spectrum and un-normalized wave function. The energy spectrum was calculated numerically using the Matlab software. The un-normalized wave function was expressed in the Hypergeometric term. The results showed that the energy spectrum increased due to the increasing minimal length parameter. The energy spectrum also increased by the increasing range of potential.

## 1. Introduction

The collective models of the nucleus are an interesting topic in nucleus area of study. The collective models which are the combination of liquid drop model and shell model [1] are used to describe the quadrupole dynamic collective in the even-even nucleus [2]. The quadrupole is the form of the deformed nucleus that corresponding to the excitation energy, and it is used to describe rotational and vibrational of the nucleus [3]. The nucleus which is considered to be in the rotation at low excitation energy is called as rigid deformed nucleus [3].

Bohr and Mottelson explained about the rotation and vibration of the nucleus in the collective models of the nucleus by using Bohr-Mottelson Hamiltonian [3]. The Bohr-Mottelson Hamiltonian has described the nucleus models with two internal variables $\beta$ and $\gamma$, and three Euler angles $(\theta, \phi, \varphi)$. The $\beta$ correspond to nucleus deformation and $\gamma$ correspond to angle symmetric [2,4]. For $\gamma = 0$ is the axially symmetric case [3,5], correspond to the prolate deformed nucleus [6] and occurs to the rigid deformed nucleus [3,5]. For $\gamma = \pi/6$ is the triaxial symmetric case and correspond to the oblate [6]. The three Euler angles $(\theta, \phi, \varphi)$ show angles of the nucleus in the collective model nucleus. In 2011, The Bohr-Mottelson Hamiltonian has been investigated by Bonatsos et al. with Davidson potential for axial symmetric [4]. The next years, the Bohr-Mottelson Hamiltonian have been solved including Eckart potential [2], Kratzer potential [7] and Hulthen plus Ring shape [8]. The methods that are used to solve Bohr-Mottelson Hamiltonian are Asymptotic Iteration Method (AIM) [8], Nikiforov-Uvarov [2] and SUSYQM [4].

Heisenberg Uncertainty Principle describes commutation relations between position and momentum operators. When Heisenberg Uncertainty Principle is influenced by quantum gravity, then this effect causes the rise of minimal observable distance in the scale Planck length [5,9], and it is well-known as General Uncertainty Principle (GUP) or minimal length [3,5,9]. The Bohr-Mottelson Hamiltonian with minimal length effect has been studied by Chabab et al. [5]. By introducing the new wave function in Bohr-

Mottelson Hamiltonian, it is reduced to the second orde differential equation then it is solvable for the case infinite square well potential with $V(\beta)=0$ [5]. In addition, Ali Mohammadi and Hassanabadi solved the Bohr-Mottelson Hamiltonian with a minimal length effect in a different way with respect to Chabab et al., they used two steps solution, where the first step of the solution is for the case with zero minimal length parameter $(\alpha_{ML}=0)$ that gives usual energy. From the first step of the solution, it was obtained the expression of Laplacian as a variable that only a function of position and usual energy spectrum [3]. The second step of the solution, the quadratic of Laplacian that is obtained from the first step is inserted into the Bohr-Mottelson Hamiltonian with a minimal length such that this equation becomes solvable for infinite square well potential [3].

In this paper, we solved Bohr-Mottelson Hamiltonian with the minimal length effect with Hulthen potential in $\beta$ function for the case of the rigid deformed nucleus as in Ref [5]. For Hulthen potential in $\beta$ function, we need Binomial expansion approximation for potential part of Bohr-Mottelson Hamiltonian. It is solved analytically using Asymptotic Iteration Method. The work is organized as follow. In section 2, the approximate solution of Bohr-Mottelson Hamiltonian with minimal length effect is briefly introduced. Asymptotic Iteration Method is reviewed in section 3. The result and discussion about the energy spectrum and un-normalized wave function are presented in section 4. Finally in section 5 conclusion is presented.

## 2. The approximate solution of Bohr-Mottelson Hamiltonian with minimal length effect

Heisenberg Uncertainty Principle is expressed by [10,11],

$$[X,P] \geq i\hbar \quad (1)$$

By considering the effect quantum gravity which is introduced as small parameter for commutation relations between position and momentum operators in (1), so (1) becomes, [10,11],

$$[X,P] \geq i\hbar\left(1+\alpha_{ML}(\Delta P)^2\right) \quad (2)$$

In (2) is the modification of commutation relations between position and momentum operators, it is called as General Uncertainty Principle (GUP). From (2), that lead to getting,

$$\hat{X}_i = \hat{x}_i \quad (3)$$

$$\hat{P}_i = \left(1+\alpha_{ML}\hat{p}^2\right)\hat{p}_i \quad (4)$$

From (3) and (4), $\alpha_{ML}$ is a minimal length parameter that has very small positive values and $X_i$ is position operator. $P_i$ and $p_i$ are momentum operators at high and low energy, respectively. The magnitude of the $P_i$ is expressed by $p$ [5]. In quantum mechanics, squared momentum operator is given by

$$\hat{P}^2 = -\hbar^2\Delta \quad (5)$$

where $\Delta$ is a Laplacian operator. We substitute (5) in (4) [2], that yield,

$$P^2 = -\frac{\hbar^2}{2B_m}(1-2\alpha_{ML}\Delta)\Delta \quad (6)$$

Momentum operator which is influenced by minimal length is shown by (6). The collective geometrical model of nucleus is expressed by [3],

$$ds^2 = \sum_{i,j} g_{ij} dx_i dx_j \quad (7)$$

In (7) is curvilinear coordinates, where $x$ is curved space and $g_{ij}$ is a metric tensor. The axial symmetry case, nucleus has three degrees of freedom: $q_1=\phi, q_2=\theta, q_3=\beta$, so the metric tensor is given by [3,5],

$$g_{ij} = \begin{pmatrix} 3\beta^2\sin^2\theta & 0 & 0 \\ 0 & 3\beta^2 & 0 \\ 0 & 0 & 1 \end{pmatrix} \quad (8)$$

where $\beta$ is a variable corresponding to nucleus deformation, $\phi$ and $\theta$ are part of Euler angles. The Laplacian operator as follows [3,5],

$$\Delta = \frac{1}{\sqrt{g}}\sum_{i,j}\frac{\partial}{\partial q_i}\sqrt{g}\,g_{ij}^{-1}\frac{\partial}{\partial q_j} \quad (9)$$

where $g$ and $g_{ij}^{-1}$ are determinant and inverse of the matrix $g_{ij}$, respectively. By using (8), it is obtained determinant of the matrix $g_{ij}$ as,

$$g = 9\beta^4\sin^2\theta \quad (10)$$

and the inverse of the matrix $g_{ij}$ as follows,

$$g_{ij}^{-1} = \begin{pmatrix} \frac{1}{3\beta^2\sin^2\theta} & 0 & 0 \\ 0 & \frac{1}{3\beta^2} & 0 \\ 0 & 0 & 1 \end{pmatrix} \quad (11)$$

By applying (10) and (11) into (9), we get Laplacian operator, is given as

$$\Delta = \left[\frac{1}{\beta^2}\frac{\partial}{\partial \beta}\beta^2\frac{\partial}{\partial \beta} + \frac{1}{3\beta^2}\left[\frac{1}{\sin\theta}\frac{\partial}{\partial \theta}\sin\theta\frac{\partial}{\partial \theta}+\frac{1}{\sin^2\theta}\frac{\partial^2}{\partial \varphi^2}\right]\right] \quad (12)$$

Hamiltonian operator is expressed by [3],

$$H = T + V(\beta) = \frac{P^2}{2B_m} + V(\beta) \quad (13)$$

where $P$ is momentum operator, $V(\beta)$ is potential energy in $\beta$ function and $B_m$ is a mass parameter. We inserted (6) and (12) in (13), so it obtained,

$$\left[-\frac{\hbar^2}{2B_m}\Delta + \frac{\alpha_{ML}\hbar^4}{B_m}\Delta^2 + V(\beta,\theta,\phi) - E\right]\psi(\beta,\theta,\phi) = 0 \quad (14)$$

Bohr-Mottelson Hamiltonian in minimal length effect is expressed (14). To solve (14) is used the new wave function [5] is given by,

$$\psi(\beta,\theta,\phi) = (1+2\alpha_{ML})\chi(\beta,\theta,\phi) \quad (15)$$

By substituted (15) and $\hbar = 1$ (natural unit) [3] in (14), it is get,

$$\left\{\Delta + \frac{2B_m(E-V(\beta,\theta,\phi))}{(1+4B_m\alpha_{ML}(E-V(\beta,\theta,\phi)))}\right\}\chi(\beta,\theta,\phi) = 0 \quad (16)$$

The solution of (16) [5] will be obtained analytically by using binomial expansion to approximation the determinator of the second term of (16), so (16) becomes,

$$\left\{\Delta + 2B_m\left[\frac{(E-V(\beta,\theta,\phi))}{(1-4B_m\alpha_{ML}(E-V(\beta,\theta,\phi)))}\right]\right\}\chi(\beta,\theta,\phi) = 0 \quad (17)$$

In (17), we have set $\alpha_{ML}^2 = 0$. The separation variable method is used to solve (17) by setting $\chi(\beta,\theta,\varphi) = R(\beta)\Theta(\theta)\Phi(\varphi)$, so we have Euler angles part of Bohr-Mottelson Hamiltonian with minimal length,

$$-\left(\frac{1}{\Phi(\varphi)}\frac{1}{\sin^2\theta}\frac{\partial^2\Phi(\varphi)}{\partial\varphi^2} + \frac{1}{\Theta(\theta)}\frac{1}{\sin\theta}\frac{\partial}{\partial\theta}\sin\theta\frac{\partial\Theta(\theta)}{\partial\theta}\right) = \lambda \quad (18)$$

and $\beta-$ part of Bohr-Mottelson Hamiltonian with minimal length,

$$\left\{\begin{array}{l}\frac{1}{\beta^2}\frac{\partial}{\partial\beta}\beta^2\frac{\partial R(\beta)}{\partial\beta} \\ +2B_m(E-V(\beta))R(\beta) \\ -8B_m^2\alpha_{ML}(E^2-2EV(\beta)+V^2(\beta))R(\beta)\end{array}\right\} = \frac{\lambda}{3\beta^2}R(\beta) \quad (19)$$

For the case potential in $\beta$ function in rigid deformed nucleus, we use $\beta-$ part of Bohr-Mottelson Hamiltonian with minimal length. $\lambda$ is constant of separation variable which is corresponding to angular momentum quantum number. By applying $R(\beta) = U(\beta)/\beta$ and $\lambda = L(L+1)$, so we have,

$$\left\{\begin{array}{l}\frac{d^2U(\beta)}{d\beta^2} - \frac{L(L+1)}{3\beta^2}U(\beta) \\ +2B_m(E-V(\beta))U(\beta) \\ -8B_m^2\alpha_{ML}(E^2-2EV(\beta)+V^2(\beta))U(\beta)\end{array}\right\} = 0 \quad (20)$$

The approximate equation of Bohr-Mottelson Hamiltonian for a $\beta-$ part in a minimal length effect for rigid deformed nucleus case is shown by (20).

## 3. Asymptotic Iteration Method

Asymptotic Iteration Method is method to solve the second order differential equation in term [12,13],

$$y_n''(t) = \lambda_0(t)y_n'(t) + s_0(t)y_n(t) \quad (21)$$

where, $\lambda_0(t) \neq 0$ and $s_0(t)$ are the coefficient of a differential equation and $n$ is a quantum number. To obtain solution, we derivative (21), so we obtain,

$$y_n^{z+1}(t) = \lambda_{z-1}(t)y_n'(t) + s_{z-1}(t)y_n(t) \quad (22)$$

where,

$$\lambda_z(t) = \lambda'_{z-1}(t) + s_{z-1}(t) + \lambda_0(t)\lambda_{z-1}(t) \quad (23)$$

$$s_z(t) = s'_{z-1}(t) + s_0(t)\lambda_{z-1}(t) \quad (24)$$

$$z = 1,2,3,... \quad (25)$$

The eigenvalue is obtained from the quantization condition which is given by,

$$\Delta_z(t) = \lambda_z(t)s_{z-1}(t) - \lambda_{z-1}(t)s_z(t) = 0 \quad (26)$$

To obtain the wave function, (21) is reduced into the formalism, as follows,

$$y_n''(t) = \left\{\begin{array}{l}2\left(\frac{at^{N+1}}{1-bt^{N+2}} - \frac{t_A+1}{t}\right)y_n'(t) \\ -\frac{wt^N}{1-bt^{N+2}}y_n(t)\end{array}\right\} \quad (27)$$

In (27) is one-dimensional Schrodinger like equation that is reduced to a hypergeometric type differential equation. The associated eigenfunction is obtained from the solution of (27), is given as,

$$y_n(t) = (-1)^n C'(N+2)^n (\sigma)_n {}_2F_1(-n, \rho+n; \sigma; bt^{N+2}) \quad (28)$$

where

$$(\sigma)_n = \frac{\Gamma(\sigma+n)}{\Gamma(\sigma)}, \quad \sigma = \frac{2t_A+N+3}{N+2}, \quad \rho = \frac{(2t_A+1)b+2a}{(N+2)b} \quad (29)$$

$C'$ is normalization constant and ${}_2F_1$ is a hypergeometric function. The un-normalized wave

function of Bohr-Mottelson Hamiltonian is obtained by using (27)-(29) [14,15].

## 4. Result and Discussion

The Hulthen potential is short range potential in physics, it is used in nuclear, particle physics, and atomic physics [16,17]. The Hulthen potential is given by [18,19],

$$V(\beta) = -V_o \frac{e^{-2\omega\beta}}{1-e^{-2\omega\beta}} \quad (30)$$

where $\omega$ and $V_o$ are a range and constant potential, respectively. For the case, atomic nucleus $V_o$ is defined as $Z\omega e^2$. The Z is an atomic number and e is a charge of the electron [20]. To get simple solution, (30) was changed in hyperbolic trigonometric term [21], is given as,

$$V(\beta) = \frac{Z\omega e^2}{2}(\coth\omega\beta - 1) \quad (31)$$

By setting the centrifugal approximate $1/\beta^2 = \omega^2/\sinh^2(\omega\beta)$ [4] and inserted (31) into (20), we obtained,

$$\frac{d^2U(\beta)}{d\beta^2} - \left[\begin{pmatrix}\frac{\omega^2 L(L+1)}{3} + 2B_m^2\alpha_{ML}V_o^2\end{pmatrix}\frac{1}{\sinh^2\omega\beta} - \begin{pmatrix}8B_m^2\alpha_{ML}V_oE + 4B_m^2\alpha_{ML}V_o^2\\-B_mV_o\end{pmatrix}\coth\omega\beta - \begin{pmatrix}2B_mE + B_mV_o - 8B_m^2\alpha_{ML}E^2\\-8B_m^2\alpha_{ML}V_oE - 4B_m^2\alpha_{ML}V_o^2\end{pmatrix}\right]U(\beta) = 0 \quad (32)$$

By setting,

$$v(v-1) = \left(\frac{\omega^2 L(L+1)}{3} + 2B_m^2\alpha_{ML}V_o^2\right) \quad (33)$$

$$2q = \left(8B_m^2\alpha_{ML}V_oE + 4B_m^2\alpha_{ML}V_o^2 - B_mV_o\right) \quad (34)$$

$$-k^2 = \begin{pmatrix}2B_mE + B_mV_o - 8B_m^2\alpha_{ML}E^2\\-8B_m^2\alpha_{ML}V_oE - 4B_m^2\alpha_{ML}V_o^2\end{pmatrix} \quad (35)$$

in (32), then we got,

$$\frac{d^2U(\beta)}{d\beta^2} - \left[\frac{v(v-1)}{\sinh^2\omega\beta} - 2q\coth\omega\beta + k^2\right]U(\beta) = 0 \quad (36)$$

The differential equation like Schrodinger equation for Manning Rosen potential [22] was shown by (36). In (36) must be reduced to hypergeometric type by using the suitable variable change $\coth(\omega\beta) = (1-2z)$, yield,

$$\left\{z(1-z)\frac{d^2U(\beta)}{dz^2} + (1-2z)\frac{dU(\beta)}{dz} + \left[v'(v'+1) - \frac{4\alpha_A^2}{4z} - \frac{4\beta_A^2}{4(1-z)}\right]U(\beta)\right\} = 0 \quad (37)$$

By setting,

$$\frac{2q-k^2}{\omega^2} = 4\alpha_A^2 \text{ and } \frac{-2q-k^2}{\omega^2} = 4\beta_A^2 \quad (38)$$

$$v'(v'+1) = \frac{v(v+1)}{3\omega^2} \quad (39)$$

where $\alpha_A$, $\beta_A$ and $v'$ are the hypergeometric parameter. In (37) is intermediate of the hypergeometric differential equation and by introducing the new wave function as,

$$U(\beta) = z^{\alpha_A}(1-z)^{\beta_A}g(z) \quad (40)$$

we obtained,

$$\left\{z(1-z)\frac{g^2(z)}{dz^2} + [(2\alpha_A+1) - (2\alpha_A+2\beta_A+2)z]\frac{g(z)}{dz} + [v'(v'+1) - (\alpha_A+\beta_A)(\alpha_A+\beta_A+1)]g(z)\right\} = 0 \quad (41)$$

By dividing (41) with $z(1-z)$, then it was reduced to AIM type equation, as

$$\left\{\frac{g^2(z)}{dz^2} + \left[\frac{(2\alpha_A+1) - (2\alpha_A+2\beta_A+2)z}{z(1-z)}\right]\frac{g(z)}{dz} + \left[\frac{v'(v'+1) - (\alpha_A+\beta_A)(\alpha_A+\beta_A+1)}{z(1-z)}\right]g(z)\right\} = 0 \quad (42)$$

By comparing (22) and (42), we had

$$\lambda_0 = \frac{-(2\alpha_A+1)}{z} + \frac{(2\beta_A+1)}{1-z} \quad (43)$$

$$s_o = \left\{\frac{(\alpha_A+\beta_A)(\alpha_A+\beta_A+1) - v'(v'+1)}{z} + \frac{(\alpha_A+\beta_A)(\alpha_A+\beta_A+1) - v'(v'+1)}{(1-z)}\right\} \quad (44)$$

The eigenvalue was obtained by using (26)-(27) and (41)-(44), is given as,

$$v'(v'+1) = (\alpha_A+\beta_A+n)(\alpha_A+\beta_A+(n+1)) \quad (45)$$

By using (38)-(39) and (45), we obtained the energy spectrum equation of Bohr-Mottelson Hamiltonian with the minimal length effect is given as,

$$E = \left\{-\frac{\omega^2}{2B_m}\left(\left(v'(v'+1) - n - \frac{1}{2}\right)^2 + \frac{\left(\frac{\delta_{\alpha_{ML}}}{2\omega^2}\right)^2}{\left(v'(v'+1) - n - \frac{1}{2}\right)^2}\right) + \xi_{E\alpha_{ML}}\right\} \quad (46)$$

with,

$$v' = \frac{1}{\omega^2}\sqrt{\frac{1}{\omega^2}\left(\frac{\omega^2 L(L+1)}{3} + 2B_m^2\alpha_{ML}V_o^2\right) + \frac{1}{4}} \quad (47)$$

$$\delta_{\alpha_{ML}} = 8B_m^2\alpha_{ML}V_oE + 4B_m^2\alpha_{ML}V_o^2 - B_mV_o \quad (48)$$

$$\xi_{E\alpha_{ML}} = 4B_m\alpha_{ML}E^2 + 4B_m\alpha_{ML}V_oE + 2B_m\alpha_{ML}V_o^2 - \frac{V_o}{2} \quad (49)$$

The energy spectrum of Bohr-Mottelson Hamiltonian with minimal length effect which was obtained by using approximate solution was expressed by (46). The $n$ is the quantum number of the nucleus. The $n=0$ was energy level of ground state band and $n=1$ was energy level of first exited. In the case $\gamma = 0$, we used $n=0$ and $n=1$ for rotation condition of the nucleus. On the other hand, for the level energy with n>1, the nucleus was in the vibration condition [5]. This paper, $B_m$ was defined as the sum of protons mass and neutrons mass. To get the energy spectrum numerically, Matlab software was used. The measurements were presented using natural units. The numerical result of energy spectrum was shown in Table 1 for $n=0$ and $L=0$ without a minimal length parameter $(\alpha_{ML}=0)$ and in the presence of minimal length parameter for some isotopes.

TABLE 1. The energy spectrum for some isotopes with $e=0.085424$ and $\omega=0.1$.

| Isotopes | E (MeV) | | |
| --- | --- | --- | --- |
| | $\alpha_{ML}=0$ | $\alpha_{ML}=0.001$ | $\alpha_{ML}=0.005$ |
| $^{102}_{44}Ru$ | -4937.10 | 22.80 | 49.37 |
| $^{106}_{46}Pd$ | -5607.80 | 25.08 | 56.08 |
| $^{114}_{48}Cd$ | -6566.90 | 29.37 | 65.67 |
| $^{132}_{54}Xe$ | -9623.80 | 43.04 | 96.24 |
| $^{138}_{56}Ba$ | -10820 | 48.39 | 108.20 |
| $^{142}_{60}Nd$ | -12781 | 57.16 | 127.81 |
| $^{152}_{62}Sm$ | -14609 | 65.33 | 146.09 |
| $^{164}_{66}Dy$ | -17862 | 79.88 | 178.62 |
| $^{192}_{76}Os$ | -27728 | 124 | 277.28 |
| $^{196}_{78}Pt$ | -29815 | 133.34 | 298.15 |

From the Table 1, it was shown that the energy spectrum was negative [17-19] without the minimal length effect, while by the presence of a minimal length effect, the energy spectrum became positive due to raising the $V^2(\beta)$ factor in (46). The isotope $^{196}_{78}Pt$ had the highest energy spectrum value with and without minimal length parameters. It was caused, $^{196}_{78}Pt$ had greater isotopes mass than the others isotopes.

The numerical result of energy spectrum was shown in Table 2 for Hulthen potential as a function of range potential $(\omega)$ for some isotopes.

TABLE 2. The energy spectrum for some isotopes with various of range potential $(\omega)$, $L=0$, and $e=0.085424$.

| Isotopes | E (MeV) | | | |
| --- | --- | --- | --- | --- |
| | $\alpha_{ML}=0.001$ | | $\alpha_{ML}=0.01$ | |
| | $\omega=0.02$ | $\omega=0.1$ | $\omega=0.02$ | $\omega=0.1$ |
| $^{102}_{44}Ru$ | 4.41 | 22.80 | 13.96 | 69.82 |
| $^{106}_{46}Pd$ | 5.01 | 25.08 | 15.86 | 79.30 |
| $^{114}_{48}Cd$ | 5.87 | 29.37 | 18.57 | 92.87 |
| $^{132}_{54}Xe$ | 8.60 | 43.04 | 27.22 | 136.10 |
| $^{138}_{56}Ba$ | 9.67 | 48.39 | 30.60 | 153.02 |
| $^{142}_{60}Nd$ | 11.43 | 57.16 | 36.15 | 180.75 |
| $^{152}_{62}Sm$ | 13.06 | 65.33 | 41.32 | 206.60 |
| $^{164}_{66}Dy$ | 15.97 | 79.88 | 50.52 | 252.60 |
| $^{192}_{76}Os$ | 24.80 | 124 | 78.43 | 392.13 |
| $^{196}_{78}Pt$ | 26.66 | 133.34 | 84.33 | 421.65 |

Table 2 showed the energy spectrum increased by the increase of the range potential $(\omega)$ value. The isotope $^{196}_{78}Pt$ had the highest energy spectrum value with the various range potential. It was caused, $^{196}_{78}Pt$ had greater isotopes mass than the others isotopes.

The numerical result of energy spectrum level for $n=0$ with the various isotopes and minimal length parameter was shown in Figure 1.

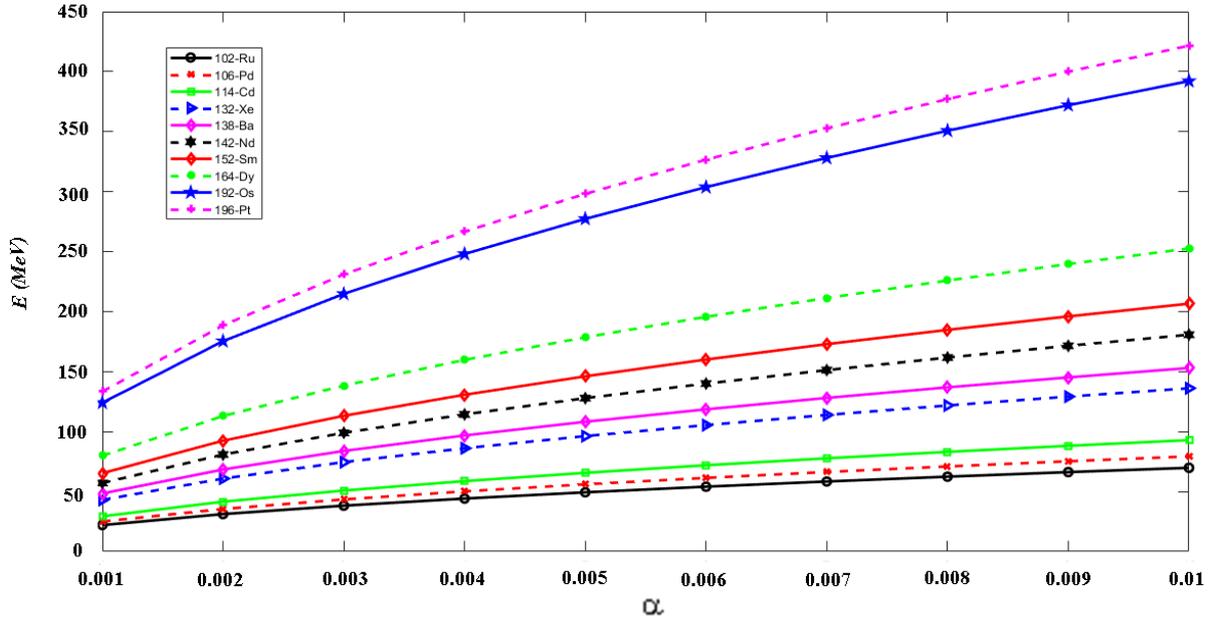

FIGURE 1. The energy spectrum for L=0 with various minimal length parameter $(\alpha_{ML})$ for some isotopes.

Figure 1 showed energy spectrum of some isotopes for n=0 with various minimal length parameter. The energy spectrum of some isotopes increases for the increasing of a minimal length parameter. For constant minimal length parameter, the energy spectrum increased, it is caused by the increasing the isotopes. The isotope which had the greatest mass had the highest energy spectrum. It was showed by $^{196}_{78}Pt$ which had highest energy spectrum. This result was in agreement with the result in Ref [3,5]. The result in ref [3,5] showed the minimal length effect increased, it caused the increasing spectra energy [3,5].

The general un-normalized wave function of Bohr-Mottelson Hamiltonian was obtained by applying (26)-(28) and (42), we get

$$U(z) = \left\{ \begin{array}{c} (-1)^n C'(1)^n (2\alpha_A+1)_n \\ {}_2F_1(-n, 2\alpha_A+2\beta_A+1+n, 2\alpha_A+1, z) \end{array} \right\} \quad (50)$$

By substituting (50) into (40) and together applying transformation variable of $\coth \omega\beta = (1-2z)$, we obtained,

$$U_0(\beta) = C'\left(\frac{1-\coth \omega\beta}{2}\right)^{\alpha_A}\left(\frac{1+\coth \omega\beta}{2}\right)^{\beta_A} \quad (51)$$

$$U_1(\beta) = \left\{ \begin{array}{c} -C'\left(\dfrac{1-\coth \omega\beta}{2}\right)^{\alpha_A}\left(\dfrac{1+\coth \omega\beta}{2}\right)^{\beta_A}(2\alpha_A+1) \\ \left[1+\dfrac{(-1)(2\alpha_A+2\beta_A+2)\left(\dfrac{1-\coth \omega\beta}{2}\right)}{(2\alpha_A+1)}\right] \end{array} \right\} \quad (52)$$

The un-normalized wave function of n=0 and n=1 are shown by (51) and (52), respectively. The wave function amplitude for some isotopes depends on the value $\alpha_A$ and $\beta_A$.

## 5. Conclusion

We investigated approximate solution of Bohr-Mottelson Hamiltonian for Hulthen potential with minimal length effect in the case of the rigid deformed nucleus. The approximate solution of Bohr-Mottelson Hamiltonian with minimal length effect was solved by using Asymptotic Iteration Method to obtain energy spectrum and un-normalized wave function. The results showed that the energy spectrum increased by the increasing of minimal length effect value for some isotopes. The energy spectrum increased due to the increasing the range potential for some isotopes. The greatest mass isotopes had the highest energy spectrum.

**Acknowledgement**
This research was partly supported by Sebelas Maret University Higher Education Project Grant Hibah Penelitian Berbasis Kompetensi 2018.

**References**
[1] W. Greiner and A.J. Maruhn,"*Nuclear Model,*" Springer, New York, 1996.

[2] L. Naderiaand H. Hassanabadi,"Bohr Hamiltonian with Eckart potential for triaxial nuclei."*The European Physical Journal Plus*, 131: 133,2016.

[3] M. Alimohammadi and H. Hassanabadi, "Alternative solution of the gamma-rigid Bohr Hamiltonian in minimal length formalism," *Nuclear Physics A*, pp. 439-449, 2017.

[4] D. Bonatsos, P. E. Georgoudis, D. Lenis, N. Minkov, and C. Quesne, "Bohr Hamiltonian with a deformation-dependent mass term for the Davidson potential, "*Physical Review C* 83, 044321,2011.

[5] M. Chabab, A. ElBatoul, A. Lahbas, and M. Oulne, "On γ-rigid regime of the Bohr–Mottelson Hamiltonian in the presence of a minimal length,"*Physics Letters B*, pp. 212-216,2016.

[6] D. Bonatsos, D. Lenis, D. Petrellis, and A.P. Terziev," Z(5) critical point symmetry for the prolate to oblate nuclear shape phase transition,"*Nuclear Physics B*, pp. 172-179, 2004.

[7] M. Alimohammadi, H. Hassanabadi and S. Zare," Investigation of Bohr-Mottelson Hamiltonian in $\gamma-$ rigid version with position dependent mass," *Nuclear Physics A*, pp. 1-13, 2017.

[8] M. Chabab, A. Lahbas, and M. Oulne, "Bohr-Hamiltonian with Hulthen plus Ring shape potential for triaxial nuclei," arXiv:1510.04525v1 [nucl-th], 2015.

[9] S Hossenfelder, "The Minimal Length And Large Extra Dimensions", *Modern Physics Letters A*, vol. 19, no. 37, pp. 2727–2744, 2004.

[10] M. Sprenger, P. Nicolini, and M. Bleicher, "Physics on the smallest scales: anintroduction to minimal lengthphenomenology", *European Journal Of Physics*, vol.33, p. 853-862. 2012.

[11] L.J. Garay,"Quantum gravity and minimum length", *International Journal of Modern Physics A*, vol.10, no.2, p. 145-165. 1994.

[12] H Ciftci, R L Hall, and N Saad," Construction of exact solutions to eigenvalue problems by the asymptotic iteration method," *J. Phys. A. Math. Gen* 36 47 11807-16, 2003.

[13] B.N. Pratiwi, A. Suparmi, C. Cari, and A.S. Husein, "Asymptotic Iteration Method for the modified Pöschl–Teller potentialand Trigonometric Scarf II non-central potentialin the Dirac equation spin symmetry,"*Pramana. J. Phys,*2017.

[14] S. Pramono, A. Suparmi, and C. Cari," Relativistic energy analysis of five-dimensional q-deformed radial Rosen-Morse potential combined with q-deformed Trigonometric Scarf Noncentral potential using Asymptotic Iteration Method,"*Adv. High Energy Phys,*2016.

[15] D. A. Nugraha, A. Suparmi, C. Cari, and B.N. Pratiwi, "Asymptotic Iteration Method for analytical solution of Klein Gordon equation for Trigonometric Pöschl-Teller potential inD-dimensions," *J. Phys. Conf. Ser.*, 2017.

[16] L. Hulthen and M. Sugawara, in Handbuch der Physik, edited by S. Flugge (Springer, Berlin,), vol. 39, 1957.

[17] Y. P. Varshni D," Eigenenergies and oscillator strengths for the Hnlthen potential,"vol.41, no.9, *Physical Review A*, 1990.

[18] Wahyulianti, A. Suparmi, C. Cari, and Kristiana N. Wea,"Construction of solvable potential partner of generalized Hulthen potential in D-dimensional Schrödinger equation,"*AIP Conference Proceedings*, pp 060011-1-6, 2017.

[19] A. Suparmi, C. Cari, and L. M. Angraini, "Bound state solution of Dirac equation for Hulthen plus Trigonometric Rosen Morse non-central potential using Romanovski polynomial,"*AIP Conference Proceedings*, pp 060011-1-6, 2015.

[20] S. M. Ikhdair,"Approximate eigenvalue and eigenfunction solutions for the generalized Hulthen potential with any angular momentum,"*Journal of Mathematical Chemistry,* vol. 42, no. 3, 2007.

[21] D. A. Dianawati, A. Suparmi, C. Cari, and Y. Mohtar, "Relativistic energy analysis for D-Dimensional Dirac equationwith Eckart plus Hulthen central potential coupled bymodified Yukawa tensor potential using Romanovskipolynomial method," *Journal of Physics: Conference Series*,776, 2016.

[22] R. A. Sari, A. Suparmi, and C. Cari," Solution of Dirac equation for Eckart potential and trigonometricManning Rosen potential using Asymptotic Iteration Method," *Chin. Phys. B*, vol. 25, no. 1, 2015.